\documentclass[aps,prd,preprint,tightenlines,superscriptaddress,nofootinbib]{revtex4}
\usepackage{amsmath,amssymb,url}
\usepackage{graphicx,tabularx}

\def\bq{\begin{equation}}
\def\eq{\end{equation}}
\def\ba{\begin{eqnarray}}
\def\ea{\end{eqnarray}}

\newcommand{\etal}{\textit{et al.}}

\newcommand{\ie}{{\sl i.e. }}
\newcommand{\eg}{{\sl e.g. }}

\newcommand{\nn}[1]{\tilde{\chi}_{#1}^0}

\newcommand{\cpi}{\tilde{\chi}_{i}^+}
\newcommand{\cmj}{\tilde{\chi}_{j}^-}
\newcommand{\msbar}{\overline{\rm MS}}
\newcommand{\drbar}{\overline{\rm DR}}

\newcommand{\GeV}{{\ensuremath\rm GeV}}

\def\ov{\overline}
\def\helas{{\sc helas}}
\def\madgraph{{\sc madgraph\;}}
\def\mgtwo{{\sc madgraph~ii\;}}
\def\madevent{{\sc madevent\;}}
\def\smadgraph{{\sc susy--madgraph\;}}
\def\wt{\widetilde}

\def\mn#1{m_{\wt{\chi}^0_#1}}
\def\mch#1{m_{\wt{\chi}^-_#1}}

\def\charginom#1{\wt{\chi}^-_#1}
\def\charginop#1{\wt{\chi}^+_#1}
\def\charginomp#1{\wt{\chi}^\mp_#1}

\def\cunitary#1{U^C_#1}
\def\nunitary#1{U^N_#1}

\def\neutralino#1{\wt{\chi}^0_#1}

\newcommand{\po}{\phantom{0}}

\begin{document}

\preprint{KEK-TH-1051, MPP-2005-152}

\date{\today}

\title{Weak boson fusion production of \\ supersymmetric
       particles at the LHC}

\author{G.~C.~Cho}
\email{cho@phys.ocha.ac.jp}
\affiliation{Ochanomizu University, Tokyo, Japan}
\author{K.~Hagiwara}
\email{kaoru.hagiwara@kek.jp}
\affiliation{Theory Division, KEK, Tsukuba, Japan}
\author{J.~Kanzaki}
\email{junichi.kanzaki@kek.jp}
\affiliation{Institute of Particle and Nuclear Studies, KEK, 
             Tsukuba, Japan}
\author{T.~Plehn}
\email{tilman.plehn@cern.ch}
\affiliation{Heisenberg Fellow, Max Planck Institute for Physics, 
             Munich, Germany \\
             and School of Physics, University of Edinburgh,
             Scotland}
\author{D.~Rainwater}
\email{rain@pas.rochester.edu}
\affiliation{Marshak Fellow, Dept.~of Physics and Astronomy,
             University of Rochester, Rochester, NY, USA}
\author{T.~Stelzer}
\email{tstelzer@uiuc.edu}
\affiliation{Dept.~of Physics, University of Illinois, 
             Urbana, IL, USA}

\begin{abstract}
We present a complete calculation of weak boson fusion production of
colorless supersymmetric particles at the LHC, using the new matrix
element generator \smadgraph.  The cross sections are small, generally
at the attobarn level, with a few notable exceptions which might
provide additional supersymmetric parameter measurements.  We discuss
in detail how to consistently define supersymmetric weak couplings to
preserve unitarity of weak gauge boson scattering amplitudes to
fermions, and derive sum rules for weak supersymmetric couplings.
\end{abstract}

\maketitle


\section{Introduction}

The matter content of the highly successful Standard Model of particle
physics is generally considered to be fully revealed after the
discovery of the top quark in 1994, although the exact mechanism of
electroweak symmetry breaking remains undetermined~\cite{EWSB}.  The
Standard Model description of spontaneous symmetry breaking is
minimal, involving only one additional complex scalar doublet.  This
introduces an unsatisfactory instability in the scalar sector of the
theory.  A theoretically more attractive scenario is that spacetime
respects the maximal extension of the Poincar\'e symmetry,
supersymmetry (SUSY)~\cite{SUSY}.  Its minimal version, the MSSM,
simultaneously provides solutions to several problems in high energy
physics and cosmology: a candidate for weakly-interacting dark matter;
possible unification of the gauge couplings at high energies; and
stability of the scalar sector which generates electroweak symmetry
breaking through renormalization group running.

SUSY must be a broken symmetry at low energy, as we do not see spin
partners of the Standard Model particles.  As a
result, the squarks, sleptons, charginos, neutralinos and gluino of
the MSSM must be massive in comparison to their Standard Model
counterparts.  Experiments such as LEP and Tevatron~\cite{run2} have
put stringent bounds on some of the SUSY partners' masses.  It will
fall to the LHC to perform a conclusive SUSY search covering masses
all the way to the TeV scale.  Real physics will, however, begin only
after a potential SUSY discovery: in particular the
strongly-interacting squarks and gluinos can be produced in large
numbers at the LHC~\cite{susy_prod,Prospino}, and their decay cascades
typically carry kinematical information about a large fraction of the
weakly interacting SUSY spectrum~\cite{cascades,bryan_spin}.  This
information can be used to narrow down different SUSY breaking
mechanisms~\cite{sfitter,pmz_unification}.

\medskip

Much MSSM and non-minimal SUSY phenomenology has been performed over
the years in preparation for LHC, nearly all of it using relatively
simple dominant $2\to 2$ processes at leading order or next-to-leading
order~\cite{susy_prod,Prospino}.  Often, these calculations involve a
number of approximations, many of which might not be sufficient for
practical applications once the collection of data
begins~\cite{catpiss}.  Examples include: consideration of spin
correlations and finite width effects in SUSY particle production and
decays; SUSY-electroweak (EW) and Yukawa interferences to some
SUSY-QCD processes; exact rather than common squark masses in the
t-channel; and additional $2\to 3$ or $2\to 4$ particle production
processes, such as the production of additional hard jets in squark
and gluino production~\cite{Plehn:2005cq}, or weak boson fusion (WBF)
production of colorless SUSY particles, the subject which we address
here.

The last topic is particularly interesting, because it may help us to
observe sleptons and weak inos at LHC.  These particles can be
extremely difficult to observe in direct production channels due to
small rates and very large backgrounds, and their appearance in squark
and gluino cascade decays can depend on the SUSY breaking scenario.
WBF is an electroweak process which naturally leads to high-$p_T$ but
very far forward taggable jets and little central jet activity.  It
can be used to observe small electroweak signal rates in a region of
phase space not very populated by QCD events.  This was applied very
successfully to heavy~\cite{WBF-H} and intermediate-mass Higgs boson
production in the Standard Model~\cite{WBF-h}, as well as weak boson
scattering~\cite{strong-WW}, the MSSM Higgs
sector~\cite{Plehn:1999nw}, and to multi--doublet Higgs
models~\cite{Alves:2003vp}.  As a consequence, there has been recent
interest in using the same technique for colorless intermediate-mass
SUSY pair production, with investigations of a few channels in limited
regions of parameter space~\cite{WBFino,WBFsf1,WBFsf2}.  The reported
results were often negative, but appeared to hint at some promising
regions of parameter space~\cite{split_susy}.  Searching for sleptons
and weak inos in these channels can provide useful information about
the SUSY Lagrangian.

We present a comprehensive calculation of supersymmetric colorless WBF
production channels at the LHC, including a broad scan over parameter
space.  We also discuss some theoretical issues regarding consistency
of electroweak gauge couplings and unitarity of weak boson scattering
to weak inos, and their relevance for practical calculations at LHC
energies.

To compute these production rates we introduce a new, supersymmetric
version of the matrix element generator \mgtwo~\cite{Maltoni:2002qb},
which properly takes into account various physics aspects which are
usually approximated in the literature, such as those listed above.
It also provides a practical means to include hard jet radiation
effects~\cite{Plehn:2005cq} and eventually match to parton shower
simulations, as well as to evaluate more complicated production
processes in LHC and linear collider phenomenology~\cite{catpiss}.  In
Sec.~\ref{sec:SMG} we discuss \smadgraph and its model assumptions,
$2\to 2$ unitarity as a test of its MSSM couplings, requirements for
MSSM electroweak couplings to maintain unitarity of weak scattering
processes at high energy, and gauge invariance of WBF processes.  We
give results for LHC WBF pair production of colorless SUSY particle in
Sec.~\ref{sec:WBF}.  We present conclusions in Sec.~\ref{sec:sum} and
a technical overview of \mgtwo and \smadgraph in the Appendix.


\section{SUSY-MadGraph, unitarity and MSSM couplings
\label{sec:SMG}}

The new matrix element generator \smadgraph is an extension of the
Standard Model \mgtwo package with the new feature of Majorana
fermions.~\footnote{Code available for download at
\url{http://pheno.physics.wisc.edu/~plehn/smadgraph} or
\url{http://www.pas.rochester.edu/~rain/smadgraph}} The implementation
of \mgtwo into the web--based event generator
\madevent~\cite{Maltoni:2002qb} is straightforward and has already
been used for the calculations published in Ref.~\cite{Plehn:2005cq}.
To include the complete MSSM particle spectrum we use a pair of model
data files describing the particle content and its interactions.
Specifically, we describe the MSSM as the minimal supersymmetric model
which conserves $R$-parity, does not contain any $CP$-violating
complex phases, and is CKM- and MNS-diagonal.  However, those
assumptions can be dropped by straightforward changes in the code.  We
do not assume any particular SUSY breaking scheme, so the MSSM
spectrum and couplings can be handed to it by any spectrum generator,
regardless of what assumptions go into constructing the spectrum.


\subsection{Setup and Tests}

The most work--intensive step in extending \mgtwo to include the MSSM
is the correct definition of all couplings in terms of the Lagrangian
parameters.  To implement the MSSM couplings, we work with the
conventions of Refs.~\cite{gunion_haber} and Ref.~\cite{TP_thesis},
and cross-check with those of Ref.~\cite{MSSMnote}.  Due to the
multiple couplings conventions in the
literature~\cite{MSSMnote,gunion_haber,Rosiek:1989rs,Kuroda:1999ks},
we employ a large number of numerical checks to ensure their
correctness.  In addition to a number of numerical comparisons to
published SUSY cross sections at $e^+e^-$~\cite{Barger:1999tn} and
$pp$ colliders~\cite{Prospino}, we also test the MSSM couplings to
electroweak bosons by checking unitarity of $2\to 2$ scattering
processes at very high energy. This not only serves to ensure that the
couplings are correct, but also reveals a general complication in the
use of spectrum generator input from the SUSY Les Houches Accord
(SLHA)~\cite{SLHA}.  SLHA is a standardized format for communicating
SUSY Lagrangian and low-energy parameters, from spectrum
generators~\cite{Allanach:2001kg,Djouadi:2002ze} and SUSY particle
widths~\cite{Muhlleitner:2003vg} to multi-purpose Monte Carlos and
next--to--leading order predictions~\cite{Prospino}.  \smadgraph uses
this convention for the SUSY spectrum input.  We discuss this
complication in subsection~\ref{sec:SLHA}.  Finally, we check some 500
$\,2\to 2$ cross sections numerically~\cite{catpiss} between
\smadgraph and the multi-purpose event generators {\sc
whizard}~\cite{whizard} and {\sc sherpa}~\cite{sherpa}.


\subsection{General {\boldmath $VV\to\tilde\chi\tilde\chi$} 
Unitarity Sum Rules\label{sec:sumrules}}

A powerful analytical and numerical check of couplings which enter the
production of two fermions in gauge boson scattering are unitarity sum
rules, which describe the scattering process in the limit of large
center-of-mass energy $E$. The general process reads:
\bq
V_1(m_1,\lambda_1) + V_2(m_2,\lambda_2)
\to 
F_1(M_1,\sigma_1) + \bar{F}_2(M_2,\sigma_2) \; .
\eq
The incoming and outgoing particle masses are $m_{1,2}$ and $M_{1,2}$,
respectively.  The incoming gauge boson polarizations are
$\lambda_{1,2}$, and the final-state fermion helicities are
$\sigma_{1,2} = L,R$.  Four types of Feynman diagrams can contribute
to the above process: $t$-channel exchange of a fermion $F_k$ of mass
$M_k$, $u$-channel exchange of a fermion $F_\ell$ with mass $M_\ell$,
annihilation to an $s$-channel vector boson $V$, or to an $s$-channel
scalar $S$.  The helicity amplitudes for these four diagrams comprise
the matrix element ${\cal M}_{\lambda_1\lambda_2}^{\sigma_1\sigma_2}$,
from which we derive the sum rules.

\smallskip

The amplitude ${\cal M}_{00}^{RR}$ can be written in terms of general
couplings $g_\pm$. For example, the interactions between two fermions
$F_1,F_2$ and a gauge boson $V$ appear in the Lagrangian as ${\cal L}
= g_\alpha^{F_1F_2V}\,\ov{F_1} \gamma^\mu P_\alpha F_2 V_\mu$.  The
left- and right-handed couplings $\alpha=\pm$ correspond to the left-
and right-handed projectors $P_\pm=(1\pm\gamma_5)/2$.  Similarly,
fermion--scalar couplings appear as ${\cal L}=
g_\pm^{F_1F_2S}\,\ov{F_1} P_\pm F_2 S$, triple--vector--boson
couplings as
${\cal L} = -i g^{V_1V_2V_3}\,
\left[
  (\partial_\mu V_{1\nu}) (V^\mu_2 V^\nu_3 - V^\nu_2 V^\mu_3)
+ (\partial_\mu V_{2\nu}) (V^\mu_3 V^\nu_1 - V^\nu_3 V^\mu_1)
+ (\partial_\mu V_{3\nu}) (V^\mu_1 V^\nu_2 - V^\nu_1 V^\mu_2)
\right]$
and vector-vector-scalar couplings as ${\cal L}= g^{V_1 V_2 S_3}\,
V_1^\mu V_{2\mu}S$.

\smallskip

We take $\hat{z}$ to be the incoming beam direction and define the
outgoing particles to lie in the $\hat{x}-\hat{z}$ plane.  The
scattering angle $\sin\theta$ describes the fraction of the final
state momentum in the $\hat{x}$ direction:
\ba
{\cal M}_{00}^{RR}
&=& \frac{2E}{m_1 m_2}\sum_{F_k}
\left[
-M_k g_+^{F_1 F_k V_1} g_-^{F_k F_2 V_2}
+ 
\left( M_1 g_-^{F_1 F_k V_1} g_-^{F_k F_2 V_2}
      +M_2 g_+^{F_1 F_k V_1} g_+^{F_k F_2 V_2}
\right)
\frac{1+\cos\theta}{2}
\right]
\nonumber \\
&+& 
\frac{2E}{m_1 m_2}\sum_{F_\ell}
\left[
-M_\ell g_+^{F_1 F_\ell V_2} g_-^{F_\ell F_2 V_1}
+ 
\left( M_1 g_-^{F_1 F_\ell V_2} g_-^{F_\ell F_2 V_1}
      +M_2 g_+^{F_1 F_\ell V_2} g_+^{F_\ell F_2 V_1}
\right)
\frac{1-\cos\theta}{2}
\right]
\nonumber \\
&+& 
\frac{E}{m_1 m_2}\sum_{V_3} g^{V_1 V_2 V_3} 
\left[ 
\left( M_1 g_-^{F_1 F_2 V_3^*} 
     - M_2 g_+^{F_1 F_2 V_3^*} 
\right)
\frac{m_1^2 - m_2^2}{m_3^2} \right]
\nonumber \\
&+&
\frac{E}{m_1 m_2}\sum_{V_3} g^{V_1 V_2 V_3} 
\left[
\left( M_1 g_-^{F_1 F_2 V_3^*} 
     + M_2 g_+^{F_1 F_2 V_3^*} 
\right)
\cos\theta
\right]
\nonumber \\
&+& 
\frac{E}{m_1 m_2}\sum_{S} g^{V_1 V_2 S} g_-^{F_1 F_2 S^*}
+ \Delta_{00}^{RR} 
\label{eq:00RR}
\ea

The remainder $\Delta_{00}^{RR}$ represents terms which do not
increase with $E$.  Each line in Eq.~(\ref{eq:00RR}) corresponds to
one class of diagrams.  The vector boson $V_3$ may be $\gamma$ or $Z$.
We note that the amplitude Eq.(2) and all equations in this section
are obtained in the unitary gauge.  A scalar field $S$, therefore,
represents both the Higgs bosons and the Goldstone bosons.  To obtain
sum rules in the Feynman gauge, remove all terms proportional to
$m^2_1-m^2_2$.

From Eq.~\ref{eq:00RR} we can derive two sum rules for $J=0$ and
$J=1$:
\ba
\underline{J=0:} \qquad &&
\sum_{F_k}
\left[
-2 M_k g_+^{F_1 F_k V_1} g_-^{F_k F_2 V_2}
+  M_1 g_-^{F_1 F_k V_1} g_-^{F_k F_2 V_2}
+  M_2 g_+^{F_1 F_k V_1} g_+^{F_k F_2 V_2}
\right]
\nonumber \\
&+&
\sum_{F_\ell}
\left[
-2 M_\ell g_+^{F_1 F_\ell V_2} g_-^{F_\ell F_2 V_1}
+  M_1 g_-^{F_1 F_\ell V_2} g_-^{F_\ell F_2 V_1}
+  M_2 g_+^{F_1 F_\ell V_2} g_+^{F_\ell F_2 V_1}
\right]
\nonumber \\
&+&
\sum_{V_3} g^{V_1 V_2 V_3} 
\left( M_1 g_-^{F_1 F_2 V_3^*} 
     - M_2 g_+^{F_1 F_2 V_3^*} 
\right)
\frac{m_1^2 - m_2^2}{m_3^2}
\nonumber \\
&+&
\sum_{S} g^{V_1 V_2 S} g_-^{F_1 F_2 S^*} = 0 \; ,
\label{eq:rr1}\\
\underline{J=1:} \qquad &&
\sum_{F_k}
\left[
  M_1 g_-^{F_1 F_k V_1} g_-^{F_k F_2 V_2}
+ M_2 g_+^{F_1 F_k V_1} g_+^{F_k F_2 V_2}
\right]
\nonumber \\
&-&
\sum_{F_\ell}
\left[
  M_1 g_-^{F_1 F_\ell V_2} g_-^{F_\ell F_2 V_1}
+ M_2 g_+^{F_1 F_\ell V_2} g_+^{F_\ell F_2 V_1}
\right]
\nonumber \\
&+&
\sum_{V_3} 
\left[
\left( M_1 g_-^{F_1 F_2 V_3^*} 
     + M_2 g_+^{F_1 F_2 V_3^*} \right) 
g^{V_1 V_2 V_3}
\right] = 0 \; .
\label{eq:rr2}
\ea
Similarly, the amplitude ${\cal M}_{00}^{LL}$ is given by 
\ba
{\cal M}_{00}^{LL}
&=& 
\frac{2E}{m_1 m_2} \sum_{F_k}
\left[
M_k g_-^{F_1 F_k V_1} g_+^{F_k F_2 V_2} 
-\left( M_1 g_+^{F_1 F_k V_1} g_+^{F_k F_2 V_2} 
      + M_2 g_-^{F_1 F_k V_1} g_-^{F_k F_2 V_2} 
 \right)
\frac{1+\cos\theta}{2}
\right]
\nonumber \\
&+& 
\frac{2E}{m_1 m_2} \sum_{F_\ell}
\left[
M_\ell g_-^{F_1 F_\ell V_2} g_+^{F_\ell F_2 V_1} 
-\left( M_1 g_+^{F_1 F_\ell V_2} g_+^{F_\ell F_2 V_1} 
      + M_2 g_-^{F_1 F_\ell V_2} g_-^{F_\ell F_2 V_1} 
 \right)
\frac{1-\cos\theta}{2}
\right]
\nonumber \\
&-& 
\frac{E}{m_1 m_2} \sum_{V_3} g^{V_1 V_2 V_3} 
\left[
\left( M_1 g_+^{F_1 F_2 V_3^*} 
     - M_2 g_-^{F_1 F_2 V_3^*} 
\right)
\frac{m_1^2 - m_2^2}{m_3^2} \right]
\nonumber \\
&-& 
\frac{E}{m_1 m_2} \sum_{V_3} g^{V_1 V_2 V_3} 
\left[
\left( M_1 g_+^{F_1 F_2 V_3^*} 
     + M_2 g_-^{F_1 F_2 V_3^*} 
\right)
\cos\theta
\right]
\nonumber \\
&-& 
\frac{E}{m_1 m_2} \sum_{S}
g^{V_1 V_2 S} g_+^{F_1 F_2 S^*}
+ \Delta_{00}^{LL} \; . 
\ea
leading to two more sum rules
\ba
\underline{J=0:}&&
\sum_{F_k}\left[
2 M_k g_-^{F_1 F_k V_1} g_+^{F_k F_2 V_2} 
- M_1 g_+^{F_1 F_k V_1} g_+^{F_k F_2 V_2} 
- M_2 g_-^{F_1 F_k V_1} g_-^{F_k F_2 V_2} 
\right]
\nonumber \\
&+&
\sum_{F_\ell}\left[
2 M_\ell g_-^{F_1 F_\ell V_2} g_+^{F_\ell F_2 V_1} 
- M_1 g_+^{F_1 F_\ell V_2} g_+^{F_\ell F_2 V_1} 
- M_2 g_-^{F_1 F_\ell V_2} g_-^{F_\ell F_2 V_1} 
\right]
\nonumber \\
&-&
\sum_{V_3} g^{V_1 V_2 V_3} 
\left( M_1 g_+^{F_1 F_2 V_3^*} 
     - M_2 g_-^{F_1 F_2 V_3^*} 
\right)
\frac{m_1^2 - m_2^2}{m_3^2}
\nonumber \\
&-&
\sum_S g^{V_1 V_2 S} g_+^{F_1 F_2 S^*} =0 \; , 
\label{eq:ll1}
\ea
\ba
\underline{J=1:}&&
\sum_{F_k}
\left[
-M_1 g_+^{F_1 F_k V_1} g_+^{F_k F_2 V_2} 
-M_2 g_-^{F_1 F_k V_1} g_-^{F_k F_2 V_2} 
\right]
\nonumber \\
&+&
\sum_{F_\ell}
\left[
  M_1 g_+^{F_1 F_\ell V_2} g_+^{F_\ell F_2 V_1} 
+ M_2 g_-^{F_1 F_\ell V_2} g_-^{F_\ell F_2 V_1} 
\right]
\nonumber \\
&+&
\sum_{V_3}
\left[
\left( - M_1 g_+^{F_1 F_2 V_3^*} 
       - M_2 g_-^{F_1 F_2 V_3^*} 
\right)
g^{V_1 V_2 V_3}
\right]
=0 \; . 
\label{eq:ll2}
\ea
The mixed--helicity amplitudes are 
\ba
{\cal M}_{00}^{LR}
&=& 
- \frac{2E^2}{m_1 m_2} \sum_{F_k} 
g_-^{F_1 F_k V_1} g_-^{F_k F_2 V_2} \sin\theta
+
\frac{2E^2}{m_1 m_2} \sum_{F_\ell} 
g_-^{F_1 F_\ell V_2} g_-^{F_\ell F_2 V_1} \sin\theta
\nonumber \\
&&-
\frac{2E^2}{m_1 m_2} \sum_{V_3}
g^{V_1 V_2 V_3} g_-^{F_1 F_2 V_3^*} \sin\theta
+ \Delta_{00}^{LR} \; , \\ 
{\cal M}_{00}^{RL}
&=& 
-
\frac{2E^2}{m_1 m_2} \sum_{F_k} 
g_+^{F_1 F_k V_1} g_+^{F_k F_2 V_2}~\sin\theta
+
\frac{2E^2}{m_1 m_2} \sum_{F_\ell} 
g_+^{F_1 F_\ell V_2} g_+^{F_\ell F_2 V_1} \sin\theta
\nonumber \\
&&-
\frac{2E^2}{m_1 m_2} \sum_{V_3}
g^{V_1 V_2 V_3} g_+^{F_1 F_2 V_3^*} \sin\theta
+ \Delta_{00}^{RL} \; ,
\ea
and lead to two sum rules 
\ba
- \sum_{F_k} g_-^{F_1 F_k V_1} g_-^{F_k F_2 V_2} 
+ \sum_{F_\ell} g_-^{F_1 F_\ell V_2} g_-^{F_\ell F_2 V_1} 
- \sum_{V_3} g^{V_1 V_2 V_3} g_-^{F_1 F_2 V_3^*} &=& 0 \; ,
\label{eq:lr} \\
- \sum_{F_k} g_+^{F_1 F_k V_1} g_+^{F_k F_2 V_2} 
+ \sum_{F_\ell} g_+^{F_1 F_\ell V_2} g_+^{F_\ell F_2 V_1} 
- \sum_{V_3} g^{V_1 V_2 V_3} g_+^{F_1 F_2 V_3^*} &=& 0 \; .
\label{eq:rl}
\ea

We note, however, that the last two sum rules are not independent from
the first four.  They can be derived from the helicity--diagonal cases
in the limits $M_{1,2}=0$.

\bigskip

To illustrate how the sum rules work we give their explicit form for
the process $W^-Z\to\charginom{1}\neutralino{1}$.  The rule for
$(R,R)$ and $J=0$ in terms of general couplings is given in
Eq.~(\ref{eq:rr1}).  If we assign $V_1=W^-, V_2=Z$ and
$F_1=\charginom{1}, F_2=\neutralino{1}\,$, we can exchange
$t$--channel neutralinos $F_k=\neutralino{k}$ and $u$--channel
charginos $F_\ell=\charginom{\ell}$.  The sum over $s$--channel
scalars does not appear in this example because the charged Higgs does
not couple to the initial--state gauge bosons.  The couplings $g_\pm$
for this case are given in the appendix.  The sum rule we have to
numerically check now becomes:
\ba
&&
\sum_{k}
\left[
-2 m_{\neutralino{k}} g_R^{\charginom{1} \neutralino{k} W} g_L^{\neutralino{k} \neutralino{1} Z}
+  m_{\charginom{1}}  g_L^{\charginom{1} \neutralino{k} W} g_L^{\neutralino{k} \neutralino{1} Z}
+  m_{\neutralino{1}} g_R^{\charginom{1} \neutralino{k} W} g_R^{\neutralino{k} \neutralino{1} Z}
\right]
\nonumber \\
&+&
\sum_{\ell}
\left[
-2 m_{\charginom{\ell}} g_R^{\charginom{1}    \charginom{\ell} Z} 
                        g_L^{\charginom{\ell} \neutralino{1}   W}
+  m_{\charginom{1}}    g_L^{\charginom{1}    \charginom{\ell} Z} 
                        g_L^{\charginom{\ell} \neutralino{1}   W}
+  m_{\neutralino{1}}   g_R^{\charginom{1}    \charginom{\ell} Z} 
                        g_R^{\charginom{\ell} \neutralino{1}   W}
\right] 
\nonumber \\
&+&
g^{WZW}
\left(
  m_{\charginom{1}} g_L^{\charginom{1} \neutralino{k} W}
- m_{\neutralino{1}} g_R^{\charginom{1} \neutralino{k} W}
\right)
\frac{m_W^2-m_Z^2}{m_W^2}
= 0
\ea

To verify these \smadgraph couplings we numerically check the set of
sum rules for the processes $W^+W^-\to\charginop{i}\charginom{j}$,
$W^-Z\to\charginom{1}\neutralino{1}$ and
$ZZ\to\neutralino{1}\neutralino{3}$.  The last process also serves as
a check for the proper description of neutralinos with negative mass
eigenvalue.  In that case we can either use a phase to re--rotate the
mass matrix onto positive eigenvalues, which means working with a
complex mixing matrix, or we can use the negative mass eigenvalue and
make use of an analytic continuation of the expression for the matrix
element~\cite{Barger:1999tn}.  The two approaches are equivalent at
leading order and respect the unitarity sum rules.

\bigskip

In addition to these sum rules, we check unitarity numerically for
amplitudes produced by {\sc susy-madgraph}.  Our test includes more
than 300 ($2\to 2$) scattering processes with the initial states
$VV$,$V\gamma$ and $VH_i$, where $V=W^\pm,Z$ and $H_i=h,H,A,H^\pm$;
and the final states $\tilde\chi^{0,\pm}_i\tilde\chi^{0,\pm,\mp}_j$,
$VH_i$, $\gamma H_i$, $H_iH_j$ and $\tilde{f}_i\tilde{f}^*_j$.  We
vary the center-of-mass energy from threshold to $10^4$~TeV, to avoid
problems with machine precision.  We require that amplitudes at most
approach a constant at high energy.  Unfortunately, $2\to 2$ unitarity
is not sufficient to check triple-Higgs (or any triple-scalar)
couplings.  Instead, we verify them by comparison with published
results~\cite{higgs_pair}.


\subsection{Unitarity and the use of SLHA input
\label{sec:SLHA}}

Let us consider a ($2\to 2$) scattering process
$VV\to\nn{i}\nn{j},\cpi\cmj$, $V=W^\pm,Z$ at high energy.  There is a
well-known gauge cancellation between $s$-, $t$- and $u$-channel
diagrams, just as in any Standard Model process $VV\to f\bar{f}$.
This limits the high-energy behavior of the amplitude to at most
approach a constant.  For the cancellation to occur, all couplings and
masses in the scattering must be exactly related by a consistent set
of electroweak parameters, driven by gauge invariance of the
Lagrangian.  Even the slightest deviation from this condition will
result in the amplitude growing as $E$ or $E^2$, depending on the
process.  This provides a powerful test of the neutralino and chargino
couplings.  However, when we check the high--energy behavior for weak
ino combinations using SLHA input from a default SUSY spectrum
generator, the test fails.  The answer lies in an inconsistent
treatment of electroweak parameters between the default assumption of
most spectrum generators, and the way collider processes are
conventionally calculated.

\smallskip

The neutralino and chargino sector is described by the mass matrices
\ba\label{eq:inomass1}
\left(
\begin{array}{cccc}
  m_{\tilde{B}} & 0  & -m_Z s_w c_\beta & \phantom{-}m_Z s_w s_\beta \\
  0 &  m_{\tilde{W}} & \phantom{-}m_Z c_w c_\beta  & -m_z c_w s_\beta \\
  -m_Z s_w c_\beta   & \phantom{-}m_Z c_w c_\beta  & 0 &  -\mu \\
  \phantom{-}m_Z s_w s_\beta & -m_Z c_w s_\beta &  -\mu & 0
       \end{array}
\right) \qquad
\left(
\begin{array}{cc}
  m_{\tilde{W}} & \sqrt{2} m_W s_\beta \\
  \sqrt{2} m_W c_\beta & \mu
\end{array}
\right)
\ea
with the bino, wino and higgsino mass parameters on the diagonal and
the gaugino--higgsino mixing masses in the off-diagonal elements.  The
neutralino and chargino mixing matrices which enter the matrix element
are computed with a given set of electroweak parameters $m_W, m_Z$ and
$s_w (=\sin\theta_w)$.  The same weak parameters also enter the matrix
element through the Standard Model gauge boson couplings.  Both sets
must be consistent to assure proper cancellation of the different
diagrams at high energy.  This is similar to Standard Model top quark
pair production, where the final state masses and the Yukawa coupling
have to be identical.  The reason for the test failure is because the
electroweak parameters used for collider calculations are generally
taken to be those at the $Z$ pole, while most spectrum generators run
the electroweak parameters up to some scale $Q$ to diagonalize the
mass matrices, where the default scale $Q$ is often the SUSY scale,
${\cal O}(1)$~TeV.  Even though the electroweak parameter differences
at these two scales may seem to be quite small, the mismatch violates
gauge invariance, \ie scattering unitarity is not preserved.

Because the renormalization group evolution implemented in spectrum
generators predicts $\drbar$ parameters at the TeV scale, our solution
to preserve unitarity is simply to extract the electroweak parameters
from the neutralino and chargino mass matrices given in the SLHA
input.  From the form of the mass matrices in Eq.~(\ref{eq:inomass1}),
extracting the effective electroweak parameters is straightforward.
If these matrices are loop-improved, this approach is equivalent to
assuming a (yet to be explored) universality~\cite{ds_tp} and
absorbing the loop corrections into the effective electroweak
parameters $m_Z^{\rm eff}, m_W^{\rm eff}, s_w^{\rm eff}$ and
$\tan\beta^{\rm eff}$.  We fix the overall coupling strength by
$G_F=1.16639\times 10^{-5}\,\GeV^{-2}$, and extract $m_Z$, $m_W$ and
$s_w^{\rm eff}$ from the mass matrices. If the weak mixing angle is
defined in the on-shell scheme, these three parameters will be
consistent also at the TeV scale.  These effective parameters are
obviously not the pole masses or proper $\msbar$ masses.  Therefore,
we should examine how large an effect running of the electroweak
parameters can have on practical calculations at colliders such as the
LHC.

\begin{figure}[t]
\includegraphics[width=0.81\textwidth]{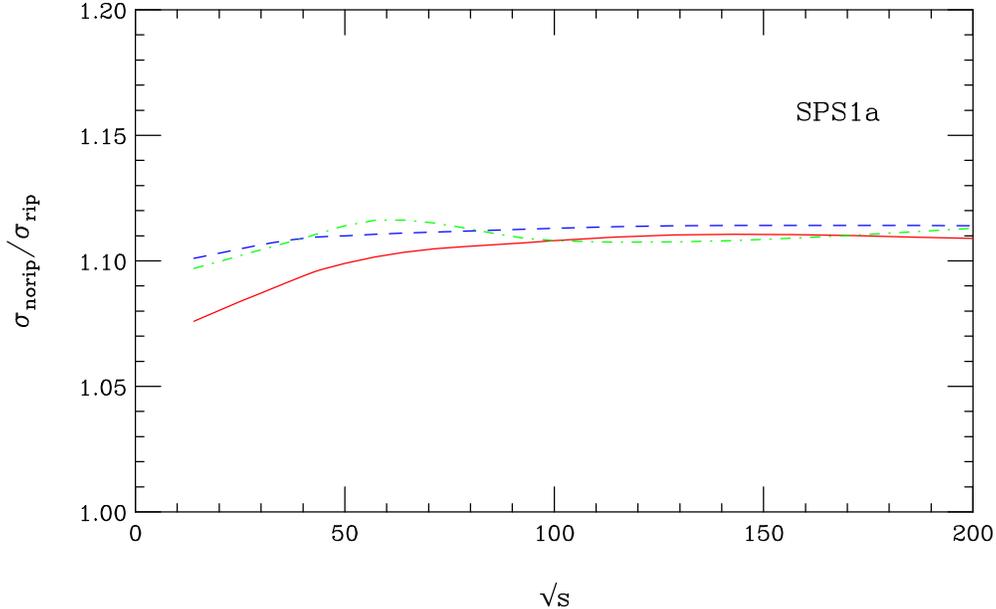}
\caption{Cross sections with the inconsistent electroweak parameters
relative to that with the consistent SLHA ripping scheme described in
the text, as function of the hadronic center-of-mass energy.  The red
(solid) curve is for $\chi^+_1\chi^-_1$ pairs, the blue (dashed) curve
for $\chi^+_1\chi^-_1$ production, and the green (dot-dashed) curve
for The cuts of Eqs.~(\ref{eq:cut1}) and (\ref{eq:cut2}) were imposed
on the two jets.}
\label{fig:xsec-rip}
\end{figure}

We quantify the impact of the different electroweak parameter schemes
in Fig.~\ref{fig:xsec-rip}, where we show the relative rates of three
particular WBF chargino and neutralino production cross sections with
and without the SLHA ripping scheme described above.  For the SUSY
spectrum, we select the generic SPS1a parameter point.  At realistic
collider energies the technical gauge-invariance violation leads to
cross section deviations at a fairly constant level of ${\cal
O}(10\%)$ for LHC through VLHC (200~TeV) energies. The flat behavior
indicates that we do not yet see unitarity violation numerically at
LHC or VLHC energies.  Nevertheless, ${\cal O}(10\%)$ corrections from
the scheme change exhaust the typical QCD error bars to WBF cross
sections~\cite{Han:1992hr}.

\begin{figure}[ht!]
\includegraphics[width=0.81\textwidth]{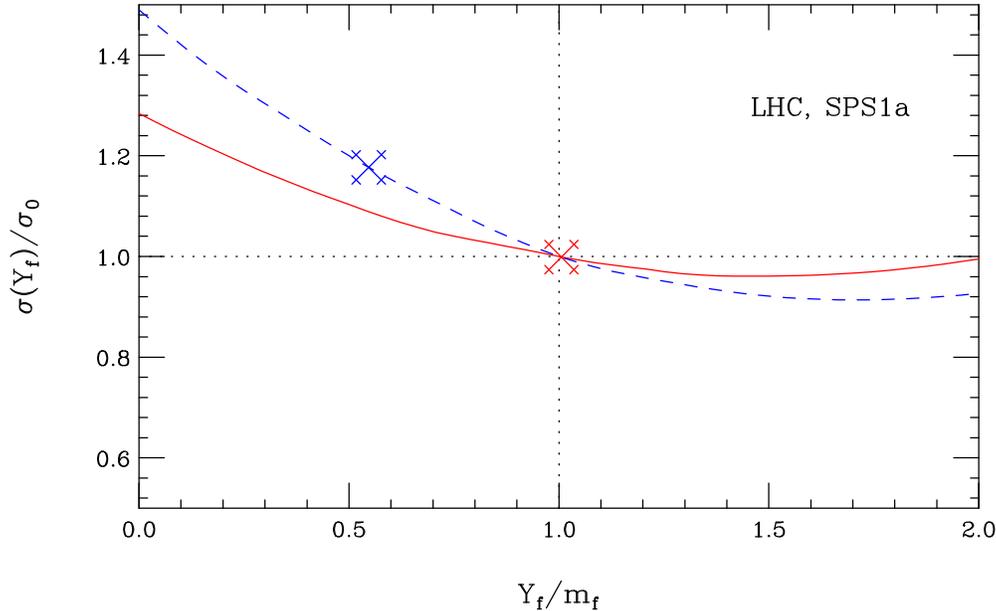}
\vspace{-2mm}
\caption{Cross section with varying Yukawa coupling relative to that 
with $\sigma_0(Y_f=m_f)$ for WBF $\tilde\tau^+_1\tilde\tau^-_1$ (red
solid) and $\tilde{b}_1\tilde{b}^*_1$ (blue dashed) production.  The
cuts of Eqs.~(\ref{eq:cut1}) and (\ref{eq:cut2}) are imposed on the
two jets.  The crosses represent the points where the Yukawa coupling
is extracted from the masses and $A_f$, as described in the text.}
\label{fig:xsec-Yuk}
\vspace{-3mm}
\end{figure}

Similar to the above--described problem with electroweak gauge
invariance and unitarity, another issue arises in the squark--Higgs
and slepton--Higgs couplings.  The symmetric scalar mass matrix is
completely defined by three entries, usually chosen as two mass
eigenvalues and the mixing angle.  They can be computed from the left-
and right-handed soft--breaking masses, the quark Yukawa coupling, the
trilinear mass parameter $A_q$ and the Higgs-sector parameters $\mu$
and $\tan\beta$.  However, for example in the
$\tilde{f}\tilde{f}^*h^0$ coupling the same parameters $m_f$, $A_f$,
$\mu$, $\tan\beta$ appear in a different combination from the
off-diagonal mass matrix entry. If we compute the matrix element for
the production process $WW\to\tilde{t}_i\tilde{t}_j$, these Lagrangian
parameters appear in the couplings to an $s$-channel Higgs, but they
also enter implicitly through the stop mixing angle (given fixed stop
masses).  Again, there is potential for a mismatch in the matrix
element.  In contrast to the electroweak parameter mismatch described
above, three-scalar couplings cannot spoil unitarity in $2\to 2$
scattering processes.  Although the mismatch breaks SUSY between the
top and stop couplings, since the SUSY violation in the three-scalar
coupling is soft, this does not reintroduce a quadratic Higgs mass
divergence at higher loop order.

There is again a way around this, but it is not as straightforward as
in the neutralino/chargino sector.  We must choose one SUSY parameter
to be extracted from the squark and slepton mass matrices, assuming
fixed mass eigenvalues.  Because the stau and stop mass matrices are
not necessarily evaluated at the same scale, the universal parameters
$\tan\beta$ and $\mu$ should not be redefined as effective parameters
to cure the mismatch in the stau and stop sectors simultaneously.  On
the other hand, the renormalization of $A_f$ has a strong impact on
the perturbative convergence of the light Higgs mass $m_h$, while
attempting to preserve SUSY by replacing the Yukawa coupling $m_f$ as
it enters the $\tilde{f}\tilde{f}^*h$ coupling also requires a shift
in the Yukawa coupling $f\bar{f}h$.  Strictly speaking, these
couplings are fixed by gauge invariance.  However, we also know that,
taking Higgs production and decay as an example, calculations using
the running $\msbar$ Yukawa couplings significantly improves the
perturbative behavior. Given these complications, and given that
keeping the mismatch corresponds to only a small shift of a
soft--breaking Lagrangian parameter, we use the usual $\drbar$
parameters $m_f$, $A_f$, $\mu$, $\tan\beta$ in the \smadgraph
couplings $\tilde{f}\tilde{f}\{h,H,A\}$.  On the other hand, for
practical purposes it will probably be preferable to use the running
Yukawa coupling in the $f\bar{f}h$ couping as well as in its
$\tilde{f}\tilde{f}^*h$ counterpart.

We investigate the practical impact of this inconsistency via the rate
change of WBF sfermion pair production (even though the rates are too
small to be observed).  For stau pairs, the rate change is typically
at the per mille level, because of the small tau Yukawa coupling.  As
another check we construct matrix elements for WBF
$\tilde{b}_1\tilde{b}^*_1$ production, even though this is a colored
final state.  The rate changes for the SPS points vary from negligible
(SPS 2,6) to almost a factor of 4 (SPS 4).  We show the cross section
rates with varying Yukawa coupling relative to those with $Y_f=m_f$ in
Fig.~\ref{fig:xsec-Yuk} for LHC $\tilde\tau^+_1\tilde\tau^-_1$ and
$\tilde{b}_1\tilde{b}^*_1$ production at SPS1a.  We see that the
Yukawa coupling extracted from the masses and from $A_f$ is indeed the
expected $\drbar$ or $\msbar$ running mass at the high scale given by
the SUSY masses.


\subsection{Interplay of WBF and non-WBF processes
\label{sec:ginv}}

\begin{figure}[ht!]
\includegraphics[width=0.78\textwidth,height=0.34\textheight]{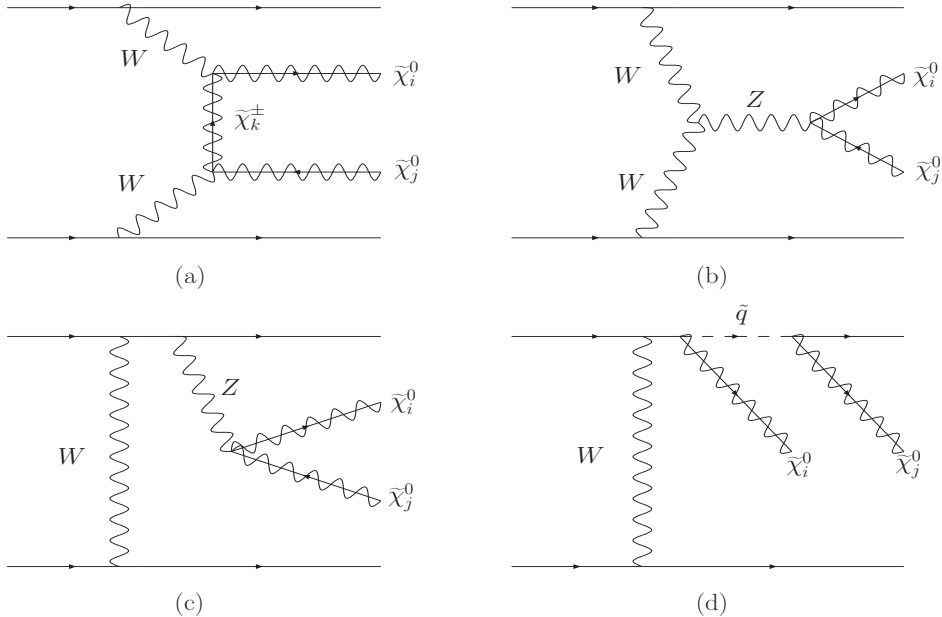}
\vspace{3mm}
\caption{Representative Feynman diagrams for WBF neutralino
pair production at a hadron collider: (a) $t$-channel chargino
exchange; (b) $s$-channel $Z$ boson (or Higgs bosons, not shown); (c)
$Z$ bremsstrahlung; (d) double--neutralino bremsstrahlung, which in
most cases is a higher-order electroweak correction to on-shell
squark-neutralino production.}
\label{fig:WBF-NiNj}
\end{figure}

As an example, we discuss neutralino pair production, with
representative Feynman diagrams shown in Fig.~\ref{fig:WBF-NiNj}.  The
$WW$-fusion component proceeds via both a $t$--channel diagram
involving a mixed gaugino and higgsino
$W^\pm\neutralino{i}\charginomp{j}$ coupling, and $s$-channel $Z$ and
Higgs boson diagrams mediated via a higgsino-induced
$Z\neutralino{i}\neutralino{j}$ coupling.  The heavy Higgs diagrams
essentially do not contribute, as the $HVV$ coupling is proportional
to $\cos(\beta-\alpha)$, which vanishes rapidly for large Higgs masses
$m_A\gtrsim 160$~GeV.  The $t$-channel diagrams of the $ZZ$-fusion
processes are induced by the higgsino content alone, and have only
Higgs bosons in the $s$-channel.

There is a class of bremsstrahlung diagrams where the incoming quarks
exchange an electroweak gauge boson, and a quark line emits a $Z$
boson which decays into a neutralino pair: Fig.~\ref{fig:WBF-NiNj}(c).
One would na\"ively estimate this contribution to be small (after the
jet cuts), but there are gauge cancellations between the diagrams
which cannot be neglected.

There also exist diagrams where a pair of incoming quarks exchange a
gauge boson in the $t$--channel, and one quark line splits to a
neutralino and a squark, which decay to neutralino plus quark,
Fig.~\ref{fig:WBF-NiNj}(d).  In other words, this is a
double-bremsstrahlung contribution, induced by the gaugino content of
the neutralino.  However, for the SPS scenarios, the squarks are
heavier than the weak inos, so these diagrams constitute higher-order
electroweak resonant neutralino-squark production.  Moreover, this
contribution does not produce the typical forward jets with central
neutralinos, and is a separate gauge set of diagrams.  Therefore, we
do not include this contribution.  Strictly speaking, for production
of light neutralinos in scenarios with relatively light squarks
(beyond the SPS benchmark points) these diagrams can be
non-negligible.

For chargino pair production, the story is much the same, although
$s$--channel $Z$ boson diagrams now also have a gaugino component.
For slepton/sneutrino production the $WW$ and $ZZ$ fusion diagrams as
well as the $Z$-bremsstrahlung are identical to the neutralino case,
but the double--bremsstrahlung diagrams do not exist.

\medskip

%

As a check of the consistent generation of amplitudes by \smadgraph,
we test the $U(1)_{EM}$ (electromagnetic) gauge invariance of WBF
production of MSSM particles.  To test electromagnetic gauge
invariance, we generate matrix elements for WBF production plus a
photon for a number of weak ino and sfermion pairs.  These are $(2\to
5)$ processes, which would be impossible to calculate without
automated tools such as {\sc susy-madgraph}.  We then set the photon
polarization vector numerically to its four-momentum, and
simultaneously set all particle widths to zero, to confirm that the
matrix element vanishes.

Although we had hoped that the EM gauge invariance of the generated
amplitudes with one additional photon emission would demonstrate the
need to include non-WBF amplitudes, mainly because the photon is
partly an electroweak ($W^3$) boson, we found that the WBF and non-WBF
amplitudes give rise to separately EM gauge invariant sets in all the
processes we studied.  It is interesting that there are further
$U(1)_{EM}$ gauge-invariant subsets, some of which are the set of all
diagrams involving a four-point $VVSS$ coupling -- that is, of two
gauge bosons and two MSSM scalars -- as well as diagrams involving two
scalars and $s$-channel Higgs propagators, as they arise from the SUSY
Lagrangian $D$-terms, which are separately gauge invariant.

Checking only the $U(1)_{EM}$ gauge-invariance of the WBF processes
is, therefore, not a verification that calculations neglecting
bremsstrahlung diagrams is gauge invariant with respect to the full
electroweak theory.  For this one has to check the exact electroweak
Ward identities for each process.  This is not as simple as replacing
the vector boson wave function with its momenta, as it is in the
photon case. Unfortunately, testing the full electroweak gauge
invariance of WBF processes is beyond the scope of this paper, as it
requires further development of automated matrix element tools.

While previous calculations~\cite{WBFino,WBFsf1,WBFsf2} are therefore
$U(1)_{EM}$ gauge-invariant, some of their results are incorrect due
to large cancellations between diagram subsets, most notably for
slepton-sneutrino production.  We quantify this in the next Section.


\section{Weak boson fusion cross section results
\label{sec:WBF}}

A characteristic of WBF is that the incoming quarks are scattered at
relatively small angles, but at typical transverse momentum of order
$\sim M_V/2$.  These scattered partons can easily be identified as
jets in CMS~\cite{cms_tdr} and ATLAS~\cite{atlas_tdr} for $p_T\gtrsim
20\;{\rm GeV}$.  We thus impose generic observability kinematic cuts
on the (parton--level) scattered quarks:
\bq\label{eq:cut1} 
p_T(j) > 20 \, {\rm GeV} \, , \qquad |\eta(j)| < 5.0 \; .
\eq
Unless the WBF process involves an internal photon propagator, which
occurs only for charged particle production, almost all the cross
section will pass these cuts.  A further characteristic of WBF is that
the two jets lie in opposite hemispheres, \ie far forward and far
backward, with large pseudorapidity separation between them.  We can
therefore impose the typical WBF separation cut
\bq\label{eq:cut2}
|\eta(j_1)-\eta(j_2)| > 4.2 \, , 
\eq
which corresponds to a pseudorapidity separation of 3 units between
the jet cone edges, using a typical LHC jet cones size of radius 0.6.
This cut typically suppresses the WBF rate by a factor 3-4.  SUSY-QCD
backgrounds which include a pair of hard jets can be a few orders of
magnitude larger than the WBF cross section before cuts. However, they
are typically suppressed by 2-3 orders of magnitude by the WBF cuts.
Normally, for a WBF analysis one imposes further constraints to ensure
the WBF-produced decay products lie between the two jets, and the two
forward jets form a large invariant mass.  These cuts usually result
in little additional loss of the signal.  Because they are not
necessary for our illustrative purposes here, we ignore such
refinement of the signal rates.  Forward-tagged jets are typically
identified with these cuts with an efficiency of $\epsilon_{\rm
tag}=0.86$, predicted by detector simulation, for low-luminosity
running.  This number might decrease somewhat for high-luminosity
running~\cite{Gianotti:2002xx}.

\smallskip

We use leading-order CTEQ6L1 parton densities~\cite{Pumplin:2002vw} in
our calculations, with factorization scale $\mu_f=(m_1+m_2)/2$, where
$m_1$ and $m_2$ are the masses of the heavy particle pair produced.
The electroweak parameters except for $G_F$ are taken from the SLHA
model files, as described in the previous section.  We investigate all
SPS points~\cite{sps}, MSSM parameter space benchmarks designed to
represent a number of canonical scenarios to align experimental
studies and phenomenology.  We consider them to be only starting
points for investigation, with no assumption that the scenarios are
more likely to be realized in nature than any other.


\subsection{Charginos and neutralinos
\label{sub:WBF-ino}}

Neutralinos and charginos can be pair-produced in WBF, for their rates
we always impose the tagging jet cuts of Eqs.(\ref{eq:cut1})
and~(\ref{eq:cut2}).  Most cross sections are at the attobarn [ab]
level, with a few exceptions.  Any cross section below 1~fb is almost
certainly going to be unobservable, unless it has a large branching
ratio to a particularly distinctive final state that can be observed
with high detection efficiency, such as multiple electrons or muons.
Even then it may have sizable backgrounds from cascade decays of
colored SUSY particles produced at QCD strength.  We comment on
possible noteworthy cases.

\smallskip

We first show the neutralino pair production cross sections in
Table~\ref{tab:SPS_nn}, which are almost universally at the
few-attobarn level.  In SPS scenarios, the lightest neutralino is
mostly bino, which reduces the coupling to weak bosons, while the
higgsinos are heavy.  The single exception to tiny neutralino pair
cross sections in all SPS scenarios is $\tilde\chi^0_2\tilde\chi^0_2$
production, where the $\tilde\chi^0_2$ is mostly wino with a sizable
higgsino fraction.  For this case, rates typically range from a tenth
of a femtobarn to half a femtobarn.  At SPS9, the $\chi^0_2$ decays
mainly to $\ell^+\ell^-\tilde\chi^0_1 (\ell=e,\mu,\tau)$, which
results in a sizable detection efficiency of due to the 4-lepton
final state with large missing transverse momentum.  Since there are
very few Standard Model backgrounds which could produce this
signature, SUSY backgrounds are the dominant source and likely small.
There is some chance this channel could be observed, although with
very low statistics unless one considers the LHC luminosity upgrade,
SLHC~\cite{Gianotti:2002xx}.

However, as we show in Table~\ref{tab:DYvWBF_inos}, the Drell-Yan (DY)
rate for $\tilde\chi^0_2\tilde\chi^0_2$ production is far
larger~\cite{Prospino}.  Unfortunately, the DY and WBF production
mechanisms are not very complementary, in the sense that both
processes involve the same gaugino--induced coupling to light-flavor
quarks and the higgsino--induced coupling to gauge bosons.  The
similar-size rate for WBF $\tilde\chi^0_1\tilde\chi^0_1$ production at
SPS9 is likely uninteresting, as the WBF $Z\to\nu\bar\nu$ rate is much
larger~\cite{Eboli:2000ze}.

\smallskip

We next show the chargino pair production cross sections in
Table~\ref{tab:SPS_xx}.  For opposite-sign production, where we do not
apply the forward tagging jet cuts as the signature is already highly
distinctive, they are at most around a femtobarn for
$\tilde\chi^+_1\tilde\chi^-_1$ at SPS1a and SPS9.  However, given that
the WBF $W^+W^-$ background is several orders of magnitude larger, we
do not expect this signal to be observable.  Moreover, WBF production
of opposite-sign charginos is again much smaller than the direct DY
production.  Much more interesting are the same--sign chargino rates,
also shown in Table~\ref{tab:SPS_xx}.  While the signal rates stay
below a femtobarn even for $\tilde\chi^+_1\tilde\chi^+_1$, the
Standard Model WBF $W^+W^+$ background is now similar in size to the
signal. Instead, the main background would more likely come from SUSY
cascades, for example from gluino pairs or squark pairs which decay to
like--sign dileptons.  Those production cross sections are typically
above $\sim 100$~pb~\cite{Prospino}, but the branching fractions which
lead to dileptons are usually tiny.  If like--sign chargino pairs were
clearly identifiable at LHC, it would provide conclusive evidence that
the neutralinos appearing in the $t$-channel are Majorana fermions,
although this would first require separate evidence that the chargino
or the neutralino candidates are in fact fermions.  We are not aware
of any other way of demonstrating this at the LHC, and extraction of
this information out of cascade decays would likely be tedious.  Of
course, a future linear collider operating in $e^-e^-$ mode would
provide definitive proof of the Majorana nature of the neutralinos.

\medskip

A complete study of signal and backgrounds for same-sign chargino
production is on the way.  The analysis breaks down into
two dominant regions of parameter space for the signal: where the
chargino decays predominantly to $W^\pm\tilde\chi^0_1$ followed by a
leptonic $W$ boson decay; and where it decays preferentially to
$\tilde\tau^\pm\nu_\tau$ and the stau ultimately decays to a lepton
and the lightest neutralino.  In both cases, the $\tilde\chi^0_1$
gives rise to significant missing transverse energy.  The final state
is then a pair of far-forward/backward high-$p_T$ jets, a pair of
same-sign central leptons (also expected to be high-$p_T$ due to the
large mass of the cascade parent), and significant missing transverse
energy.  There would also be very little additional jet activity, due
to the electroweak nature of the scattering process, as mentioned in
the Introduction.  The efficiency for such a signal would be extremely
high.  The Standard Model WBF $W^\pm W^\pm$ background is also of
${\cal O}(1)$~fb~\cite{strong-WW}, but should display markedly
different lepton kinematics.  SUSY backgrounds would arise primarily
from QCD squark and gluino production, which are often collectively at
the tens to hundreds of picobarn level.  However, only a small
fraction of these will give high-$p_T$ forward jets from the cascade
decay to jets plus charginos, and much of the sample will have
additional high-$p_T$ jets which can be vetoed.  The lepton kinematics
will again be significantly different, as the typical relatively light
charginos (compared to the squarks and gluinos) will yield much
higher-$p_T$ leptons.

\medskip

Finally, we show the mixed chargino-neutralino production cross
sections in Tables~\ref{tab:SPS_x+n} and~\ref{tab:SPS_x-n}.  These are
also at the attobarn level, the only exception being
$\chi^\pm_1\chi^0_2$ production, with a typical rate of ${\cal
O}(1)$~fb.  However, this should be compared to the DY rate of many
hundreds of femtobarn as shown in Table~\ref{tab:DYvWBF_inos}.

\begin{table}[h!]
\begin{tabular}{|c||>{\po}l<{\po}|>{\po}l<{\po}|>{\po}l<{\po}|>{\po}l<{\po}
                   |>{\po}l<{\po}|>{\po}l<{\po}|>{\po}l<{\po}|>{\po}l<{\po}
                   |>{\po}l<{\po}|>{\po}l<{\po}|}
\hline
\, SPS \, 
&   1a  &   1b  &    2  &    3  &    4  &    5  &    6  &    7  &    8  &    9  \\
\hline
$\chi^0_1\chi^0_1$ 
& 0.003 & 0     & 0     & 0     & 0     & 0.001 & 0     & 0.001 & 0     & 0.46  \\
$\chi^0_1\chi^0_2$ 
& 0.018 & 0.001 & 0.002 & 0.001 & 0.004 & 0.003 & 0.003 & 0.008 & 0.003 & 0     \\
$\chi^0_1\chi^0_3$ 
& 0.002 & 0     & 0     & 0     & 0.001 & 0     & 0.001 & 0.003 & 0.001 & 0.002 \\
$\chi^0_1\chi^0_4$ 
& 0.002 & 0     & 0     & 0     & 0.001 & 0     & 0.001 & 0.001 & 0.001 & 0.002 \\
$\chi^0_2\chi^0_2$ 
& 0.52  & 0.10  & 0.24  & 0.10  & 0.26  & 0.29  & 0.039 & 0.057 & 0.15  & 0     \\
$\chi^0_2\chi^0_2$ 
& 0.049 & 0.008 & 0.009 & 0.008 & 0.026 & 0.008 & 0.009 & 0.017 & 0.016 & 0     \\
$\chi^0_2\chi^0_4$ 
& 0.065 & 0.011 & 0.011 & 0.011 & 0.034 & 0.009 & 0.023 & 0.045 & 0.022 & 0     \\
$\chi^0_3\chi^0_3$ 
& 0.006 & 0.001 & 0.001 & 0.001 & 0.004 & 0.001 & 0.004 & 0.009 & 0.003 & 0     \\
$\chi^0_3\chi^0_4$ 
& 0.008 & 0.001 & 0.001 & 0.001 & 0.004 & 0.001 & 0.005 & 0.013 & 0.003 & 0     \\
$\chi^0_4\chi^0_4$ 
& 0.007 & 0.001 & 0.001 & 0.001 & 0.004 & 0.001 & 0.008 & 0.020 & 0.003 & 0     \\
\hline
\end{tabular}\centering
\caption{Cross sections~[fb] for WBF neutralino pair production at LHC, 
for all MSSM benchmark SPS points, using the kinematic cuts of
Eqs.~(\ref{eq:cut1}) and (\ref{eq:cut2}).  Cross sections are shown to
two significant digits or rounded to the nearest attobarn, and those
smaller than half an attobarn are shown as zero.}
\label{tab:SPS_nn}
\end{table}
\begin{table}[h!]
\begin{tabular}{|c||>{\po}l<{\po}|>{\po}l<{\po}|>{\po}l<{\po}|>{\po}l<{\po}
                   |>{\po}l<{\po}|>{\po}l<{\po}|>{\po}l<{\po}|>{\po}l<{\po}
                   |>{\po}l<{\po}|>{\po}l<{\po}|}
\hline
\, SPS \, & 1a & 1b & 2 & 3 & 4 & 5 & 6 & 7 & 8 & 9 \\
\hline
$\chi^+_1\chi^-_1$ 
& 1.6   & 0.26  & 0.63  & 0.27  & 0.74  & 0.77  & 0.13  & 0.23  & 0.42  & 1.3   \\
$\chi^+_1\chi^-_2$ 
& 0.056 & 0.010 & 0.011 & 0.010 & 0.029 & 0.010 & 0.015 & 0.028 & 0.019 & 0.002 \\
$\chi^+_2\chi^-_2$ 
& 0.035 & 0.007 & 0.004 & 0.006 & 0.020 & 0.003 & 0.030 & 0.068 & 0.015 & 0     \\
\hline
$\chi^+_1\chi^+_1$ 
& 0.93  & 0.22  & 0.48  & 0.23  & 0.51  & 0.57  & 0.067 & 0.077 & 0.31  & 0.88  \\
$\chi^+_1\chi^+_2$ 
& 0.13  & 0.022 & 0.028 & 0.022 & 0.070 & 0.015 & 0.072 & 0.14  & 0.049 & 0.002 \\
$\chi^+_2\chi^+_2$ 
& 0.001 & 0     & 0     & 0     & 0.001 & 0     & 0.011 & 0.032 & 0.001 & 0     \\
\hline
$\chi^-_1\chi^-_1$ 
& 0.28  & 0.056 & 0.13  & 0.058 & 0.14  & 0.16  & 0.017 & 0.020 & 0.083 & 0.25  \\
$\chi^-_1\chi^-_2$ 
& 0.040 & 0.006 & 0.005 & 0.006 & 0.021 & 0.005 & 0.018 & 0.036 & 0.014 & 0.001 \\
$\chi^-_2\chi^-_2$ 
& 0     & 0     & 0     & 0     & 0     & 0     & 0.002 & 0.007 & 0     & 0     \\
\hline
\end{tabular}\centering
\caption{Cross sections~[fb] for WBF opposite-sign and same-sign chargino
pair production at LHC, for all MSSM benchmark SPS points, using the
kinematic cuts of Eqs.~(\ref{eq:cut1}) and (\ref{eq:cut2}) for
opposite-sign charginos only (see text).  Cross sections are shown to
two significant digits or rounded to the nearest attobarn, and those
smaller than half an attobarn are shown as zero.}
\label{tab:SPS_xx}
\end{table}
\begin{table}[h!]
\begin{tabular}{|c||>{\po}l<{\po}|>{\po}l<{\po}|>{\po}l<{\po}|>{\po}l<{\po}
                   |>{\po}l<{\po}|>{\po}l<{\po}|>{\po}l<{\po}|>{\po}l<{\po}
                   |>{\po}l<{\po}|>{\po}l<{\po}|}
\hline
\, SPS \, & 1a & 1b & 2 & 3 & 4 & 5 & 6 & 7 & 8 & 9 \\
\hline
$\chi^+_1\chi^0_1$ 
& 0.015 & 0.001 & 0.001 & 0.001 & 0.004 & 0.002 & 0.003 & 0.010 & 0.003 & 0.53  \\
$\chi^+_1\chi^0_2$ 
& 0.64  & 0.10  & 0.26  & 0.11  & 0.30  & 0.31  & 0.044 & 0.074 & 0.17  & 0     \\
$\chi^+_1\chi^0_3$ 
& 0.042 & 0.007 & 0.007 & 0.007 & 0.021 & 0.007 & 0.008 & 0.018 & 0.013 & 0.001 \\
$\chi^+_1\chi^0_4$ 
& 0.044 & 0.007 & 0.008 & 0.007 & 0.024 & 0.006 & 0.012 & 0.022 & 0.015 & 0.001 \\
\hline
$\chi^+_2\chi^0_1$ 
& 0.003 & 0     & 0     & 0     & 0.001 & 0     & 0.001 & 0.001 & 0.001 & 0.001 \\
$\chi^+_2\chi^0_2$ 
& 0.029 & 0.005 & 0.005 & 0.005 & 0.015 & 0.004 & 0.010 & 0.021 & 0.010 & 0     \\
$\chi^+_2\chi^0_3$ 
& 0.007 & 0.001 & 0.001 & 0.001 & 0.004 & 0.001 & 0.005 & 0.011 & 0.003 & 0     \\
$\chi^+_2\chi^0_4$ 
& 0.008 & 0.001 & 0.001 & 0.001 & 0.004 & 0.001 & 0.008 & 0.021 & 0.003 & 0     \\
\hline
\end{tabular}\centering
\caption{Cross sections~[fb] for WBF positive-sign chargino plus 
neutralino pair production at LHC, for all MSSM benchmark SPS points,
using the kinematic cuts of Eqs.~(\ref{eq:cut1}) and (\ref{eq:cut2}).
Cross sections are shown to two significant digits or rounded to the
nearest attobarn, and those smaller than half an attobarn are shown as
zero.}
\label{tab:SPS_x+n}
\end{table}
\begin{table}[h!]
\begin{tabular}{|c||>{\po}l<{\po}|>{\po}l<{\po}|>{\po}l<{\po}|>{\po}l<{\po}
                   |>{\po}l<{\po}|>{\po}l<{\po}|>{\po}l<{\po}|>{\po}l<{\po}
                   |>{\po}l<{\po}|>{\po}l<{\po}|}
\hline
\, SPS \, & 1a & 1b & 2 & 3 & 4 & 5 & 6 & 7 & 8 & 9 \\
\hline
$\chi^-_1\chi^0_1$ 
& 0.009 & 0.001 & 0.001 & 0.001 & 0.002 & 0.001 & 0.002 & 0.006 & 0.002 & 0.32  \\
$\chi^-_1\chi^0_2$ 
& 0.39  & 0.059 & 0.15  & 0.062 & 0.18  & 0.19  & 0.026 & 0.044 & 0.097 & 0     \\
$\chi^-_1\chi^0_3$ 
& 0.027 & 0.004 & 0.005 & 0.004 & 0.014 & 0.005 & 0.006 & 0.012 & 0.009 & 0.001 \\
$\chi^-_1\chi^0_4$ 
& 0.027 & 0.005 & 0.005 & 0.004 & 0.015 & 0.004 & 0.007 & 0.014 & 0.009 & 0.001 \\
\hline
$\chi^-_2\chi^0_1$ 
& 0.002 & 0     & 0     & 0     & 0.001 & 0     & 0     & 0.001 & 0.001 & 0.001 \\
$\chi^-_2\chi^0_2$ 
& 0.018 & 0.003 & 0.003 & 0.003 & 0.010 & 0.003 & 0.006 & 0.013 & 0.006 & 0     \\
$\chi^-_2\chi^0_3$ 
& 0.004 & 0.001 & 0     & 0.001 & 0.002 & 0.001 & 0.003 & 0.007 & 0.002 & 0     \\
$\chi^-_2\chi^0_4$ 
& 0.005 & 0.001 & 0     & 0.001 & 0.003 & 0     & 0.005 & 0.012 & 0.002 & 0     \\
\hline
\end{tabular}\centering
\caption{Cross sections~[fb] for WBF negative-sign chargino plus 
neutralino pair production at LHC, for all MSSM benchmark SPS points,
using the kinematic cuts of Eqs.~(\ref{eq:cut1}) and (\ref{eq:cut2}).
Cross sections are shown to two significant digits or rounded to the
nearest attobarn, and those smaller than half an attobarn are shown as
zero.}
\label{tab:SPS_x-n}
\end{table}
\begin{table}[ht!]
\begin{tabular}{|c||>{\po}l<{\po}|>{\po}l<{\po}|>{\po}l<{\po}|>{\po}l<{\po}
                   |>{\po}l<{\po}|}
\hline
& \multicolumn{2}{c|}{SPS 1a} & \multicolumn{2}{c|}{SPS8} \\
\hline
      & \;\,DY\;\, & WBF & \;\,DY\;\, & WBF \\
\hline
$\chi^0_2\chi^0_2$ & \po25.4 & 0.52 & \po\po2.90 & 0.15 \\
$\chi^+_1\chi^-_1$ & 705     & 1.6  & 212        & 0.42 \\
$\chi^+_1\chi^0_2$ & 828     & 0.64 & 276        & 0.17 \\
$\chi^-_1\chi^0_2$ & 474     & 1.26 & 142        & 0.31 \\
\hline
\end{tabular}\centering
\caption{Drell-Yan (DY) and WBF pair production cross sections~[fb] 
for two representative MSSM benchmark SPS points. The DY results were
calculated with Prospino2.0~\protect\cite{Prospino} at NLO.}
\label{tab:DYvWBF_inos}
\end{table}
%


\subsection{Sleptons and sneutrinos
\label{sub:WBF-sf}}

The second type of supersymmetric WBF processes is the production of
charged sleptons and sneutrinos, again with the tagging jet cuts of
Eqs.~(\ref{eq:cut1}) and~(\ref{eq:cut2}).  Smuon cross sections are
identical to those for selectrons, as their masses are identical, so
we omit them in our tables.

Sneutrino pair production rates, shown in Table~\ref{tab:SPS_vv}, are
at the attobarn level and universally too small to be observed.  The
charged slepton cross sections and the mixed cases
(cf. Tables~\ref{tab:SPS_ll} and~\ref{tab:SPS_lv}) show similarly
limited promise (in contradiction with Refs.~\cite{WBFsf1,WBFsf2}).
At least part of this discrepancy is due to large cancellations
between the pure-WBF diagrams and bremsstrahlung, where the incoming
quarks scatter off each other via a weak boson, and the sparticle pair
arises from $Z/W$-Bremsstrahlung off a quark line.  The cancellation
is practically nil for the neutral current ($ZZ$ fusion) in slepton
pair production, as well as sneutrino pairs, but a little more than a
factor 3 in the charged current ($WW$ fusion), which dominates,
leaving an overall cancellation of about a factor 3.  The cancellation
is much larger for mixed sneutrino-slepton pairs, about a factor 30.
Refs.~\cite{WBFsf1,WBFsf2} considered only the WBF diagrams and so did
not notice the cancellations.  We emphasize that these cancellations
reflect electroweak gauge invariance, which we discussed previously.

A useful comparison is to note that the Drell-Yan rates for selectron
and stau pairs, shown in Table~\ref{tab:DYvWBF_slep}, are in the few
tens of fb range for many SPS points.  One might have hoped to use
just the WBF sneutrino pair signals, which would often give missing
energy plus forward-tagged jet signals, but because of the ultra-low
signal rates, this will again be swamped by WBF $Z\to\nu\bar\nu$ boson
production~\cite{Eboli:2000ze}.

\begin{table}[h!]
\begin{tabular}{|c||>{\po}l<{\po}|>{\po}l<{\po}|>{\po}l<{\po}|>{\po}l<{\po}
                   |>{\po}l<{\po}|>{\po}l<{\po}|>{\po}l<{\po}|>{\po}l<{\po}
                   |>{\po}l<{\po}|>{\po}l<{\po}|}
\hline
\, SPS \, & 1a & 1b & 2 & 3 & 4 & 5 & 6 & 7 & 8 & 9 \\
\hline
$\tilde\nu_e\tilde\nu^*_e$
&\po0.028 &\po0.004 & 0     &\po0.007 & 0.001 &\po0.011 &\po0.003 &\po0.011 &\po0.003 & 0.002 \\
$\tilde\nu_\tau\tilde\nu^*_\tau$
&\po0.027 &\po0.004 & 0     &\po0.007 & 0.002 &\po0.011 &\po0.003 &\po0.011 &\po0.003 & 0.002 \\
\hline
\end{tabular}\centering
\caption{Cross sections~[fb] for WBF sneutrino pair production at LHC, 
for all MSSM benchmark SPS points.  Cross sections are shown to two
significant digits or rounded to the nearest attobarn, and those
smaller than half an attobarn are shown as zero.}
\label{tab:SPS_vv}
\end{table}
\begin{table}[ht!]
\begin{tabular}{|c||>{\po}l<{\po}|>{\po}l<{\po}|>{\po}l<{\po}|>{\po}l<{\po}
                   |>{\po}l<{\po}|>{\po}l<{\po}|>{\po}l<{\po}|>{\po}l<{\po}
                   |>{\po}l<{\po}|>{\po}l<{\po}|}
\hline
\, SPS \, & 1a & 1b & 2 & 3 & 4 & 5 & 6 & 7 & 8 & 9 \\
\hline
$\tilde{e}_L^+\tilde{e}_L^-$
& 0.052 & 0.008 & 0 & 0.014 & 0.003 & 0.022 & 0.006 & 0.022 & 0.007 & 0.004 \\
$\tilde{e}_R^+\tilde{e}_R^-$
& 0.045 & 0.006 & 0 & 0.021 & 0.001 & 0.017 & 0.008 & 0.069 & 0.023 & 0.001 \\
\hline
$\tilde\tau_1^+\tilde\tau_1^-$
& 0.053 & 0.016 & 0 & 0.023 & 0.006 & 0.019 & 0.008 & 0.075 & 0.025 & 0.002 \\
$\tilde\tau_1^+\tilde\tau_2^-$
& 0.007 & 0.003 & 0 & 0.001 & 0.001 & 0.002 & 0     & 0.002 & 0     & 0.001 \\
$\tilde\tau_2^+\tilde\tau_2^-$
& 0.043 & 0.006 & 0 & 0.014 & 0.003 & 0.020 & 0.006 & 0.021 & 0.006 & 0.002 \\
\hline
\end{tabular}\centering
\caption{Cross sections~[fb] for WBF charged slepton pair production
at LHC, for all MSSM benchmark SPS points.  Cross sections are shown
to two significant digits or rounded to the nearest attobarn, and
those smaller than half an attobarn are shown as zero.}
\label{tab:SPS_ll}
\end{table}
\begin{table}[h!]
\begin{tabular}{|c||>{\po}l<{\po}|>{\po}l<{\po}|>{\po}l<{\po}|>{\po}l<{\po}
                   |>{\po}l<{\po}|>{\po}l<{\po}|>{\po}l<{\po}|>{\po}l<{\po}
                   |>{\po}l<{\po}|>{\po}l<{\po}|}
\hline
\, SPS \, & 1a & 1b & 2 & 3 & 4 & 5 & 6 & 7 & 8 & 9 \\
\hline
$\tilde{e}_L^+\tilde\nu_e$
& 0.026 & 0.003 & 0 & 0.006 & 0.001 & 0.010 & 0.002 & 0.010 & 0.003 & 0.002 \\
$\tilde\tau_1^+\tilde\nu_\tau$
& 0.005 & 0.003 & 0 & 0.001 & 0.001 & 0.001 & 0     & 0.002 & 0     & 0.001 \\
$\tilde\tau_2^+\tilde\nu_\tau$
& 0.023 & 0.003 & 0 & 0.006 & 0.001 & 0.009 & 0.002 & 0.010 & 0.003 & 0.001 \\
\hline
\end{tabular}\centering
\caption{Cross sections~[fb] for WBF charged slepton plus sneutrino
pair production at LHC, for all MSSM benchmark SPS points.  Cross
sections are shown to two significant digits or rounded to the nearest
attobarn, and those smaller than half an attobarn are shown as zero.
We neglect showing the charge-conjugate processes as the cross
sections are trivially small.}
\label{tab:SPS_lv}
\end{table}
\begin{table}[ht!]
\begin{tabular}{|c||>{\po}l<{\po}|>{\po}l<{\po}|>{\po}l<{\po}|>{\po}l<{\po}
                   |>{\po}l<{\po}|}
\hline
& \multicolumn{2}{c|}{SPS 1a} & \multicolumn{2}{c|}{SPS8} \\
\hline
\, process \, & \;\,DY\;\, & WBF & \;\,DY\;\, & WBF \\
\hline
$\tilde{e}_L^+\tilde{e}_L^-$
& 22.5 & 0.052 & \po2.49 & 0.007 \\
$\tilde{e}_R^+\tilde{e}_R^-$
& 29.0 & 0.045 & 14.3    & 0.023 \\
\hline
$\tilde\tau_1^+\tilde\tau_1^-$
& 34.4 & 0.053 & 16.0    & 0.025 \\
$\tilde\tau_2^+\tilde\tau_2^-$
& 18.3 & 0.043 & \po2.40 & 0.006 \\
\hline
\end{tabular}\centering
\caption{Cross sections [fb] for Drell-Yan (DY) v.\ WBF slepton pair 
production at LHC, for two representative MSSM benchmark SPS points.
Cross sections are shown to two significant digits or rounded to the
nearest attobarn, and those smaller than half an attobarn are shown as
zero.  The DY results were calculated with
Prospino~\protect\cite{Prospino} at NLO.}
\label{tab:DYvWBF_slep}
\end{table}
%


\section{Conclusions
\label{sec:sum}}

We have presented a thorough investigation of weak boson fusion
production of colorless SUSY particles at the LHC.  We find the cross
sections to be almost universally unobservably small for all SPS
scenarios, usually at the few-attobarn level, especially in the case
of sleptons and sneutrinos.  The smallness of the cross sections is
partly due to large cancellations which take place at the amplitude
level between WBF-type diagrams and bremsstrahlung diagrams.

\smallskip

There remain two or three exceptions, where the rates are potentially
interesting: $\tilde\chi^0_2\tilde\chi^0_2$ production, which would
give a highly distinctive four-lepton final state in many MSSM
scenarios; and same-sign chargino production
$\tilde\chi^\pm_i\tilde\chi^\pm_i$.  The latter case is especially
intriguing, because it can constitute definitive proof already at LHC
that the neutralinos in the $t$-channel are Majorana fermions.  This
is a crucial test for a candidate SUSY discovery.  It might also
provide information on the relative hierarchy of the wino and higgsino
mass parameters.  We sketched how such an analysis would proceed,
identifying the dominant final state channels as a function of MSSM
parameterization and outlining the major relevant backgrounds.

\medskip

To perform these calculations, we developed the matrix element
generator \smadgraph which includes the complete MSSM with $R$-parity
and without additional $CP$ violation, and does not assume any
particular SUSY breaking scheme.  It relies on the input from any SUSY
spectrum generator in the SLHA format.  We tested the MSSM
implementation by comparing with the literature for all known $2\to 2$
collider scattering processes, and additionally via unitarity of
scattering amplitudes at high energy and $U(1)_{EM}$ gauge invariance.
We furthermore derived analytical sum rules for neutralino and
chargino electroweak gauge couplings from unitarity in all modes of
$VV\to\tilde\chi\tilde\chi$ scattering.

We identified an issue of electroweak gauge invariance using SLHA
input during development, on account of the unitarity checks: the
default output of SUSY spectrum generators, combined with the default
use of electroweak parameters for collider calculations, forms an
inconsistent set of electroweak parameters.  This misalignment can
lead to sizable deviations in physical cross sections, relative to
the known level of QCD perturbative uncertainty in the overall rates.
\smadgraph uses electroweak parameters taken from the neutralino and 
chargino mixing matrices to form a consistent set.


\begin{acknowledgments}
We are grateful for the support of the DESY theory group, the Madison
Phenomenology Institute and the MPI for Physics in Munich while we
were writing and testing \smadgraph.  We also thank Wolfgang Kilian,
J\"urgen Reuter, Steffen Schumann and Frank Krauss. Moreover, we would
like to thank Graham Kribs and his Ultra-Mini Workshop on BSM Physics
as well as the Aspen Center for Physics, where part of this work was
finalized, for their hospitality.

This research was supported in part by the U.S. Department of Energy
under grant No. DE-FG02-91ER40685 (DR) and the U.S. National Science
Foundation under award No. 0426272 (TS), as well as the Grant-in-Aid
for Scientific Research of Ministry of Education, Culture, Sports,
Science and Technology, Japan (No.13740149 and No.16028204 for GCC,
No.17540281 for KH and JK).
\end{acknowledgments}


\appendix


\section{MadGraph II technical details
\label{sec:MG2}}

\madgraph~\cite{Stelzer:1994ta} is a package which generates matrix 
elements for a user-input scattering process at a specified order in
$\alpha_s$ and $\alpha$.  The Fortran output it provides calls the
\helas~\cite{Murayama:1992gi} library of helicity amplitude
subroutines, which can calculate any possible dimension-4 Lagrangian
term.  It is extremely versatile, allowing as many external particles
as one has computing power to handle, and allowing the user to require
certain intermediate states while restricting others.  It maintains
all helicity correlations throughout.

The original version had the Standard Model explicitly incorporated into the
Fortran code.  To facilitate additions beyond the Standard Model such as SUSY,
\madgraph was rewritten such that it could read model information
from an input file, and also handle Majorana fermions.  These new
features are incorporated into the release of \mgtwo, designed to work
with \madevent~\cite{Maltoni:2002qb}, and briefly described below.
The new version also now provides sub-amplitudes in leading-$N_c$
color flows that allow it to be interfaced to standard parton-shower
Monte Carlo programs such as {\sc pythia}~\cite{pythia}, {\sc
herwig}~\cite{herwig} or {\sc sherpa}~\cite{sherpa}.

The existence of Majorana fermions in SUSY required another
significant modification to the code.  \madgraph's use of \helas\
requires a well-defined continuous fermion flow for the calculation of
amplitudes and their proper interference.  This continuous flow is not
automatically satisfied by processes that involve Majorana fermions.
Therefore, for a given process \mgtwo defines a continuous fermion
flow~\cite{Denner:1992me}, chosen randomly from the two possible
direction for any fermion line.  For every fermion, \madgraph also
defines a charge conjugate fermion with the opposite flow.  Using this
complete set of fermions, and requiring continuity of the fictitious
fermion flow, \mgtwo is able to generate all of the appropriate
diagrams, with the proper interference structure.  In this scheme,
conflicting fermion flows (``clashing arrows'') are avoided by
choosing a fermion flow randomly from the two possible directions,
charge-conjugating the external wavefunction on one end of the fermion
flow to match the chosen direction, and calculating the
charge-conjugated vertex at the point of clashing arrows.

\smallskip

The Standard Model Lagrangian in \mgtwo is specified in two files,
defining the particle content and the dimension-4 interactions,
respectively. The user must set the defined couplings in a separate
Fortran routine, with values to be passed to the matrix elements by
common block. The first file, particles.dat, defines all particles in
the model.  The next few lines show the format of the
space-delineated file.

\begin{verbatim}
d       d~        F        S      ZERO  ZERO    T   d    1
t       t~        F        S      TMASS TWIDTH  T   t    6
e-      e+        F        S      ZERO  ZERO    S   e    11
g       g         V        C      ZERO  ZERO    O   _    21
w-      w+        V        W      WMASS WWIDTH  S   W    -24
h       h         S        D      HMASS HWIDTH  S   h    25
p uu~dd~ss~cc~g
\end{verbatim}

The first two entries define the characters to be used for the
particle and antiparticle respectively.  Notice in the case of a gluon
it is the same symbol. The third entry gives the spin of the particle,
choices are F for spin 1/2 fermions, V for spin 1 vectors and S for
spin 0 scalars. The fourth entry defines the type of line in the
Feynman diagram, SCWD correspond to solid, curly, wavy, and dashed
respectively.  The next two entries are text strings representing the
particles mass and width. The seventh entry gives the color
information, valid options are STO, representing color singlet,
triplet or octet.  The eighth entry is the text to be displayed on the
Feynman diagram labeling the line, and the last entry is a number
representing the particles ID to be used in interfacing with the QCD
Les Houches Accord. In this format the complete set of Standard Model
particles can be specified with just 17 lines.

At the LHC it is particularly useful to generate a set of subprocesses
by summing over groups of particles, \eg partons in the proton, or
final--state jets.  The last line above defines the symbol p to
represent the sum over the light quarks and gluon. There can not be
any spaces between the particles that are to be summed over.

\bigskip

The second file necessary to define a model is interactions.dat.  As
the name implies this file contains information about the allowed
interactions.  \mgtwo currently allows for 3- and 4-particle vertices
to be implemented using the space-delimited format shown below.

\begin{verbatim}
d d g GG QCD
g g g G  QCD
g g g g G G QCD QCD
d d a  GAD QED
d u w- GWF QED
u d w+ GWF QED
\end{verbatim}

The first line shows the vertex for a $d\bar{d}g$ QCD interaction.
The first three entries are the particles appearing at the dimension-4
vertex.  The ordering convention for fermion-fermion-boson
interactions is first the incoming fermion, followed by the outgoing
fermion, and the outgoing boson last.  This is also illustrated in the
interactions of the $W^+$ and $W^-$.  The fourth entry contains the
coupling strength for the interaction as it appears in the Fortran
file, and the last entry specifies the type of coupling. This final
string can be used to limit the diagrams to a certain order if
$\alpha$ or $\alpha_s$.  The four--gluon vertex illustrates a
four-particle interaction.  The first four entries define the
particles involved in the interaction, the product of the fifth and
sixth entries gives the couplings for the interaction, and the last
two entries specify the couplings involved.

This is all the information required to define an interaction.  The
color structure of the vertex is inferred based on the color of the
particles, and the Lorentz structure is inferred based on their spins.
This format is compact, and the complete Standard Model (using
diagonal CKM and MNS matrices) is specified with only 58 lines.

\bigskip

\smadgraph is the \mgtwo package with two data files (as described 
above) which completely specify the MSSM Lagrangian.  Moreover, it
includes a Fortran routine which reads in the MSSM parameters from an
SLHA files, and another which calculates the MSSM couplings.  The
input file may be generated by any MSSM spectrum generator.  If one
wants to include particle decays in the matrix elements, their widths
must be specified in the SLHA file, and are read in automatically by
the code.  The couplings are assumed to conserve $CP$, but are written
in the include file in such a way as to facilitate later inclusion of
$CP$-violation, if the user chooses to add it.  Chargino and
neutralino mixing matrices are defined to be real.  Negative
neutralino masses appearing in the propagators do not introduce any
errors for the Majorana scheme we implemented.


\section{Neutralino and Chargino Feynman Rules
\label{sec:app_feynman}}

In this appendix we give the explicit Feynman rules and couplings
needed to evaluate the example sum rule for the process
$W^-Z\to\charginom{1}\neutralino{1}$ given as an example at the end of
Section~\ref{sec:sumrules}.  Unfortunately, there are (at least) two
sets of conventions for the chargino and neutralino mixing matrices
present in the literature, namely Refs.~\cite{gunion_haber,TP_thesis}
and Ref.~\cite{MSSMnote}.  We give all formulas in both conventions.


\subsection{Neutralino and Chargino Mixing}

The chargino mass matrix according to Ref.~\cite{MSSMnote} reads
\ba\label{eq:charginomass2}
M_C &=& 
        \left( \begin{array}{cc}
        M_2 & \sqrt{2} m_W c_\beta \\
        & \\
        \sqrt{2} m_W s_\beta & \mu 
        \end{array} \right). 
\ea
and can be diagonalized using two unitary matrices
${\cunitary{R}}^\dagger M_C \cunitary{L} = {\rm diag}(\mch{j})$.

The unitary mixing matrices $U^C_{L,R}$ can be expressed in terms of
two mixing angles. As long as these mixing matrices are real, one of
the mass eigenvalues can in principle be negative. In that case a
phase can be introduced in one of the mixing matrices (\eg $U^C_R$),
or (as long as $CP$ is conserved) we can perform the matrix element
calculation with negative mass eigenvalues, simply by analytically
continuing the expressions for the matrix element.  Note that in
\helas we have to provide positive mass values for the spinor 
calculations.  The analytic continuation holds only for the matrix
element.

\smallskip

The chargino mixing matrices can easily be translated into the
conventions used in Refs.~\cite{gunion_haber, TP_thesis}, with the
chargino mixing matrix Eq.~(\ref{eq:inomass1}).  In that case we use
the unitary matrices $U$ and $V$ to diagonalize the transpose of the
chargino mixing matrix defined in Eq.~(\ref{eq:charginomass2})
$U^*{\cal M}_C^T V^{-1}$.  The translation rule becomes:
\bq
U_L^C = U^\dagger \; ,\qquad \qquad U_R^C = V^T \; .
\eq

The neutralino mass matrix in both sets of conventions is identical,
so we just copy the expression from Eq.~(\ref{eq:inomass1}):
\ba
M_N &=& 
        \left( \begin{array}{cccc}
        M_{\tilde{B}} & 0 & -m_Z s_w c_\beta &  m_Z s_w s_\beta \\
        0 & M_{\tilde{W}} &  m_Z c_w c_\beta & -m_Z c_w s_\beta \\
        -m_Z s_w c_\beta  &  m_Z c_w c_\beta & 0 & -\mu \\
         m_Z s_w s_\beta  & -m_Z c_w s_\beta & -\mu & 0 \\
        \end{array} \right) \; .
\ea
It is diagonalized as ${\nunitary{R}}^\dagger M_N \nunitary{L} = {\rm
diag}(\mn{j})\,$.  Because the neutralinos are Majorana fermions, the
mass matrix $M_N$ is symmetric.  Hence, up to a matrix of phase
factors $P$, two unitary matrices $\nunitary{L}$ and $\nunitary{R}$
can be chosen as $\nunitary{L} = U_N P^*$ and $\nunitary{R} = U_N^*
P$.  Again, if $CP$ is conserved and we are willing to work with
negative neutralino mass eigenvalues using analytic continuation of
the matrix element expression, we can choose $P=1$.  Alternatively, we
can absorb the phase of the mass eigenvalue into a then-complex mixing
matrix.

\smallskip

Again, we can translate the neutralino mixing matrices in the
conventions of Refs.~\cite{gunion_haber,TP_thesis}.  There, the mass
matrix in the bino--wino basis is diagonalized by a unitary
transformation $N^*{\cal M}_N N^{-1}$.  The mixing matrices are now
related by:
\bq
U_L^N \equiv U^N = N^\dagger \; ,\qquad \qquad
U_R^N = (U_L^N)^* = N^T \; .
\eq
%


\subsection{Couplings for interactions with gauge bosons}

Using the above definitions, we write out the couplings of gauge
bosons to neutralinos and charginos for both set of conventions
defined above.  The symbol $U$ without any superscript refers to the
chargino mixing matrix according to
Refs.~\cite{gunion_haber,TP_thesis}.  Note that the fermion order in
\helas and in the sum rules is $g_{L,R}^{f_{\rm out}f_{\rm in}V}$,
while in the \smadgraph file interactions.dat it is $g_{R,L}^{f_{\rm
in}f_{\rm out}V}$, where $f_{\rm in, out}$ are not necessarily both
particle (as opposed to antiparticle).  Instead, they are the particle
or antiparticle for the rule as read directly off a Feynman diagram of
the interaction vertex.  This is important to remember in constructing
chargino rules for \smadgraph.  The mixed chargino--neutralino--$W$
couplings are:

\ba
g_L^{\neutralino{i} \charginom{j} W} 
        &=& \left( g_L^{\charginom{j} \neutralino{i} W} \right)^* 
        = -g \left[ 
        \left( \nunitary{L} \right)_{2i}^* 
        \left( \cunitary{L} \right)_{1j}
        + 
        \frac{1}{\sqrt{2}} 
        \left( \nunitary{L} \right)_{3i}^* 
        \left( \cunitary{L} \right)_{2j}
        \right] 
\\
        & & \phantom{ \left( g_R^{\charginom{j} \neutralino{i} W} \right)^* }
        = -g \left[ 
        N_{i2} \;
        U_{j1}
        + 
        \frac{1}{\sqrt{2}} 
        N_{i3} \; 
        U_{j2}
        \right] \; ,
\nonumber \\ 
g_R^{\neutralino{i} \charginom{j} W} 
        &=& \left( g_R^{\charginom{j} \neutralino{i} W} \right)^* 
        = -g \left[ 
        \left( \nunitary{R} \right)_{2i}^* 
        \left( \cunitary{R} \right)_{1j}
        -
        \frac{1}{\sqrt{2}} 
        \left( \nunitary{R} \right)_{4i}^* 
        \left( \cunitary{R} \right)_{2j}
        \right] 
\\
        & & \phantom{ \left( g_L^{\charginom{j} \neutralino{i} W} \right)^* }
        = -g \left[ 
        N_{i2}^* \; 
        V_{j1}
        -
        \frac{1}{\sqrt{2}} 
        N_{i4}^* \;
        V_{j2}
        \right] \; .
\nonumber
\ea
All neutralino and chargino mixing matrices are defined above and $g$
is the usual weak gauge coupling.  The chargino couplings to a photon
or a $Z$ boson are proportional to the gauge couplings $g_Z=g/c_w$,
with the usual weak mixing angle $c_w=\cos\theta_w$ and
$s_w=\sin\theta_w$:
\ba
g_L^{\charginom{i} \charginom{j} Z} 
        &=& g_Z \left[ 
         \left( \cunitary{L} \right)^*_{1i} 
        \left( \cunitary{L} \right)_{1j} 
        + \frac{1}{2} 
        \left( \cunitary{L} \right)^*_{2i} 
        \left( \cunitary{L} \right)_{2j} 
        - s_w^2 \delta_{ij}
        \right]
\\
        &=& g_Z \left[ 
        U^*_{i1} \; 
        U_{j1} 
        + \frac{1}{2} 
        U^*_{i2} \;
        U_{j2} 
        - s_w^2 \delta_{ij}
        \right] \; ,
\nonumber \\ 
g_R^{\charginom{i} \charginom{j} Z} 
        &=& g_Z \left[ 
        \left( \cunitary{R} \right)^*_{1i} 
        \left( \cunitary{R} \right)_{1j} 
        + \frac{1}{2} 
        \left( \cunitary{R} \right)^*_{2i} 
        \left( \cunitary{R} \right)_{2j} 
        - s_w^2 \delta_{ij}
        \right]
\\
        &=& g_Z \left[ 
        V^*_{i1} \; 
        V_{j1} 
        + \frac{1}{2} 
        V^*_{i2} \;
        V_{j2} 
        - s_w^2 \delta_{ij}
        \right] \; ,
\nonumber \\ 
g^{\charginom{i} \charginom{i} A}_L  &=& 
        g^{\charginom{i} \charginom{i} A}_R  = e \; .
\ea
Finally, the neutralino coupling to the $Z$ is:
\ba\label{eq:NiNjZ_L}
g_L^{\neutralino{i} \neutralino{j} Z}
        &=& -\frac{1}{4} g_Z \left[ 
        \left( \nunitary{L} \right)^*_{3i}
        \left( \nunitary{L} \right)_{3j}
        - 
        \left( \nunitary{L} \right)^*_{4i}
        \left( \nunitary{L} \right)_{4j}
        \right] \; ,
\\
\phantom{g_L^{\neutralino{i} \neutralino{j} Z}}
        &=& -\frac{1}{4} g_Z
            \left[ 
                N_{i3} \; N^*_{j3}
              - N_{i4} \; N^*_{j4}
            \right] \; ,
\nonumber \\ \label{eq:NiNjZ_R}
g_R^{\neutralino{i} \neutralino{j} Z}
        &=& -g_L^{\neutralino{j} \neutralino{i} Z} \; .
\ea
Due to the Majorana nature of the neutralinos, the indices $i,j$
should be taken as $i<j$, and there is an additional factor of 2 for
$i=j$ in Eqs.~(\ref{eq:NiNjZ_L}) and (\ref{eq:NiNjZ_R}).
For the sum rules in Feynman gauge we need the Goldstone couplings to
$W,Z$, neutralinos and charginos:
\bq
g^{W^- Z\omega^+} = -g^{W^+ Z\omega^-} = -ig_Z m_W s_w^2
\eq
\ba
g_L^{\neutralino{i} \charginom{j} \omega^+}
&=&
\left( g_R^{\charginom{j} \neutralino{i} \omega^+} \right)^*
\\\nonumber
&=&
- \frac{\displaystyle i}{\sqrt{\displaystyle 2}} g \left[
                             \left( \nunitary{R} \right)^*_{2i} 
                             \left( \cunitary{L} \right)_{2j}
                            -\sqrt{2} 
                             \left( \nunitary{R} \right)^*_{3i} 
                             \left( \cunitary{L} \right)_{1j}
                            +\frac{\displaystyle s_w}{\displaystyle c_w}
                             \left( \nunitary{R} \right)^*_{1i} 
                             \left( \cunitary{L} \right)_{2j}
                       \right] c_\beta
\\\nonumber
&=&
- \frac{\displaystyle i}{\sqrt{\displaystyle 2}} g \left[
                             N^*_{i2} 
                             U_{j2}
                            -\sqrt{2} 
                             N^*_{i3} 
                             U_{j1}
                            +\frac{\displaystyle s_w}{\displaystyle c_w}
                             N^*_{i1} 
                             U_{j2}
                       \right] c_\beta
\ea
\ba
g_R^{\neutralino{i} \charginom{j} \omega^+}
&=&
\left( g_L^{\charginom{j} \neutralino{i} \omega^+} \right)^*
\\\nonumber
&=&
- \frac{\displaystyle i}{\displaystyle \sqrt{2}} g \left[
                             \left( \nunitary{R} \right)_{2i} 
                             \left( \cunitary{R} \right)_{2j}
                            +\sqrt{2}
                             \left( \nunitary{L} \right)^*_{4i} 
                             \left( \cunitary{R} \right)_{1j}
                            +\frac{\displaystyle s_w}{\displaystyle c_w}
                             \left( \nunitary{R} \right)_{1i} 
                             \left( \cunitary{R} \right)_{2j}
\right] s_\beta
\\\nonumber
&=&
- \frac{\displaystyle i}{\displaystyle \sqrt{2}} g \left[
                             N_{i2} 
                             V_{j2}
                            +\sqrt{2}
                             N^*_{i4} 
                             V_{j1}
                            +\frac{\displaystyle s_w}{\displaystyle c_w}
                             N_{i1} 
                             V_{j2}
\right] s_\beta
\ea
%


\end{document}